\documentclass[twoside, epsfig, 12pt]{article}
\usepackage{graphicx}

\input oejv_cl.sty

\begin{document}

\OEJVhead{June 2008}
\OEJVtitle{The 2008 February superoutburst of V452 Cas}
\OEJVauth{Christopher Lloyd$^1$, Roger Pickard$^2$, Jeremy Shears$^3$, 
Ian Miller$^4$,}
\OEJVauth{David Boyd$^5$ and Steve Brady$^6$}
\OEJVinst{Open University, Milton Keynes MK7 6AA, UK {\tt C.Lloyd@open.ac.uk}}
\OEJVinst{3 The Birches, Shobdon, Leominster, Herefordshire, HR6 9NG, UK \\
{\tt rdp@astronomy.freeserve.co.uk}}
\OEJVinst{Pemberton, School Lane, Bunbury, Tarporley, Cheshire, CW6 9NR, UK
{\tt bunburyobservatory@hotmail.com}}
\OEJVinst{Furzehill House, Ilston, Swansea, SA2 7LE, UK {\tt
furzehillobservatory@hotmail.com}}
\OEJVinst{5 Silver Lane, West Challow, Wantage, Oxon, OX12 9TX, UK
{\tt drsboyd@dsl.pipex.com}}
\OEJVinst{5 Melba Drive, Hudson, NH 03051, USA {\tt sbrady10@verizon.net}}

\OEJVabstract{Observations of the 2008 February outburst of V452 Cas show that 
the profile, duration and magnitude at maximum were very similar to the previous
superoutburst in 2007 September. Low-amplitude variations consistent with 
previously observed superhumps were also seen.}

\begintext

V452 Cas is a poorly observed UGSU class of dwarf nova with a range of $V \sim 
15.3 - 18.5$ and was thought to have rare outbursts. Shears et al. (2008) have 
recently reported an observing campaign on this system and have revealed that 
it is much more active than previously thought, although the faintness of the 
outbursts means that they can be easily missed. V452 Cas was first identified 
as a dwarf nova by Richter (1969) and limited observations by Bruch et al. 
(1987) caught one outburst. 
Spectroscopic confirmation of the dwarf nova classification was provided by 
Liu and Hu (2000) when the star was in quiescence at $V=18.6$. Systematic visual
monitoring did not begin until 1993 but most of these early observations had a
limiting magnitude brighter than 15.0 so were unlikely to detect an outburst. 
Nevertheless, the first visual discovery was made later that year.

\begin{table}[b]\centering
\caption{Equipment used}
\begin{tabular}{lll} \hline
Observer&	Telescope&	CCD\\
Miller&	0.35 m SCT&	Starlight Xpress SXVF-H16 \\
Pickard& 0.30 m SCT&	Starlight Xpress SXVF-H9\\
Shears&	0.28 m SCT&	Starlight Xpress SXV-M7\\
\hline
\end{tabular}
\end{table}

Shears et al. have searched the data in the AAVSO International Database and 
identified eight probable outbursts between 1989 and 2005. Most of these are 
poorly observed, with sometimes just the discovery observation. However, in 
two outbursts times series observations revealed superhumps. These were first 
seen during the 1999 November outburst by Vanmunster and Fried (1999) who 
found the superhump period, $P_{sh} = 0.0891(4)$ d, and then during the 2000 
September 
outburst by the Kyoto group (Kato 2000), although no period was reported. During
the 2007 September superoutburst Shears et al. found a progression from early 
superhumps
with $P_{sh} = 0.08943(7)$ d to late superhumps with $P_{sh} = 0.08870(2)$ d,
at about 3 or 4 days into the outburst.

During an intensive observing campaign between 2005 and 2008 Shears et al. 
discovered a further 23 outbursts. Most of these are short with durations of 
$< 4$ days and are never brighter than 16.0C while 7 have durations $>8$ days 
and 
are also brighter than 16.0C. The shorter outbursts are identified as normal 
outbursts while the longer and brighter ones are superoutbursts. This division 
is consistent with what is known about the outbursts with superhumps. The 1999 
September superoutburst lasted for 12 days and reached 14.7vis. Unfortunately 
there are insufficient observations of the 2000 September superoutburst to
gauge its duration but it 
did reach 15.1vis. The 2007 September superoutburst was observed extensively 
by Shears et al. and lasted at least 11 days, but was possibly a little 
fainter than the others, reaching 15.3C at maximum. 

From the timings of the superoutbursts Shears et al. were able to derive a 
superoutburst period and the ephemeris has been used to help predict when these 
might be 
seen. The initial ephemeris suggested that one would occur in the middle 
of 2008 January (Shears and Lloyd, 2007), but the two outbursts discovered 
around this time, on 2008 January 4 and February 1 (Miller, 2008), both turned 
out to be normal outbursts. The superoutburst was finally seen about a month 
later but was not well observed due to a combination of poor visibility since it 
was 
close to solar conjunction, and poor weather. The equipment used is given in 
Table 1 and the sequence was the same as that used by Shears et al., provided 
by the AAVSO (Henden and Sumner 2002) and given in Appendix 1. The unfiltered
observations were reduced as $V$ and all the data are given in Appendix 2.

Two unfiltered observations on the two nights before the outburst put the 
system fainter than 17.2C. On 2008 February 28 it was seen in outburst at 
16.0C and four days later it was slightly brighter at 15.7C. The following 
night a single observations gave $V=15.63$ and on the same night a long time 
series gave $V \sim 15.7$. The second, shorter time series five days later put 
the star in decline at $V \sim 16.2$. 
\begin{figure}[hb]\centering
\includegraphics[width=8cm]{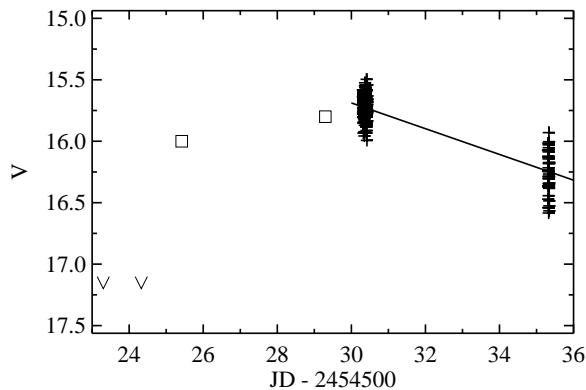}
\caption{Light curve of the 2008 February superoutburst of V452 Cas showing the
two upper limits prior to the outburst and the general profile of the outburst.
Single unfiltered observations are shown as open squares and the time series $V$ 
data as crosses. The other single $V$ measurement is obscured by the first time
series.}
\label{fig:outburst}
\end{figure}
The profile of the outburst is shown in Figure~1 
and it closely mirrors the 2007 superoutburst described in detail by Shears et 
al. From the profile the diagnostic features of the superoutburst are the 
duration, in this case $>10$ days and the magnitude at maximum, which is $V \sim
15.7$. The observations near the middle of the outburst probably 
correspond to those at the end of the plateau phase in the 2007 superoutburst 
and the second time series probably occurs two or three days prior to the 
final fade to quiescence. The decline rate of 0.10 mag/day over the second 
half of the outburst is very similar to the decline rate of the 2007 
superoutburst over the same period. The best alignment of the two outbursts 
suggests that the 2007 superoutburst was discovered a day after it went into 
outburst. The onset of the 2008 superoutburst is well defined by the negative
observations, but there is a few days uncertainty for the 2007 superoutburst so
it is quite possible that it was discovered one day after the outburst began.

The long time series was taken under poor conditions 
and the errors are larger than expected, so to improve the visibility of any
variation a 5-point median filter has been applied and the independent values
are shown in Figure 2. There is a suggestion of a cyclical 
variation and that has been fitted by a second order Fourier series with the
late-superhump period from Shears et al. It is clear that the first harmonic is
very strong and gives the impression that the period is half the expected value.
Shears et al. did not see any variation similar to this but day six of the 2007
superoutburst, probably corresponding to $JD = 2454349$ was not observed. The
light curve from two days later does show a likely increase in the first
harmonic but not to the extent seen here. The amplitude at this stage of the
2007 superoutburst was 0.2 magnitudes and that is similar to the range seen
here.

The outburst characteristics of V452 Cas are now much better understood than 
previously but it is still an object worthy of continued attention. The normal 
outbursts are faint and short and can be easily missed but the superoutbursts 
are seen much more reliably. Shears et al. have identified all the 
superoutbursts since 2005 so it would be useful to extend this complete record 
to study their longer term behaviour. The next two superoutbursts are expected 
in 2008 June/July and November/December when the star is well placed for
northern hemisphere observers.
\begin{figure}[bh]\centering
\includegraphics[width=8cm]{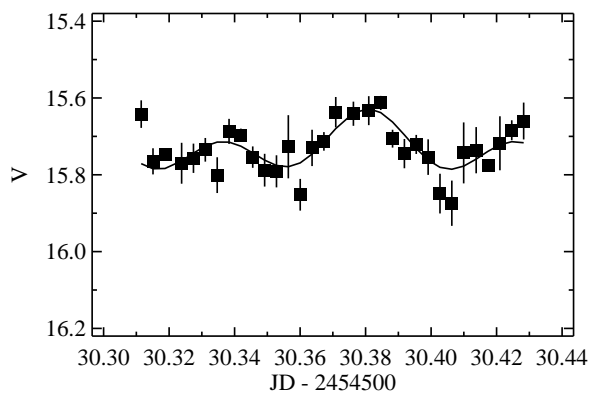}
\caption{Light curve of the long run showing the independent 5-point median
filtered data with the Fourier fit fixed at the late-superhump period, 
$P_{sh} = 0.08870$ d.}
\label{fig:outburstN1}
\end{figure}

\subsection*{Acknowledgements} This research has made use of the VizieR 
catalogue access tool and the SIMBAD database, operated at CDS, Strasbourg, 
France, and NASA's Astrophysics Data System.

\references

Bruch A., Fischer F.J., Wilmsen U., 1987, 
Astron. Astrophys. Suppl. Ser., {\bf70}, 481-516
\OEJVbibcode{1987A\&AS...70..481B}

Henden A., Sumner B., 2002, AAVSO finding chart 020301 and sequence F or FR chart
\OEJVlink{http://www.aavso.org/observing/charts/}

Kato T., 2000, vsnet-campaign 510 
\footnote{The link to vsnet-campaign frequently times out but is generally
available, in contrast to vsnet-alert messages which are not available}
\\
\OEJVlink{http://www.kusastro.kyoto-u.ac.jp/vsnet/Mail/vsnet-campaign/msg00510.html}

Liu W., Hu J.Y., 2000, Astrophys. J. Suppl. Ser., {\bf128}, 387-401
\OEJVbibcode{2000ApJS..128..387L}

Miller I., 2008, baavss-alert 1509 and 1555 \\
\OEJVlink{http://tech.groups.yahoo.com/group/baavss-alert/message/1509} \\
\OEJVlink{http://tech.groups.yahoo.com/group/baavss-alert/message/1555}

Richter G.A., 1969, Mitt. Ver\"{a}nderliche Sterne, {\bf 5}, 69-72 
\OEJVbibcode{1969MitVS...5...69R} \\
\OEJVlink{http://www.stw.tu-ilmenau.de/observatory/observatory\_3\_1.html}
\footnote{The link to MVS documents has changed recently from \\
http://www.stw.tu-ilmenau.de/science/pub/MVS/texts/MVS... and the link to the
on-line document from the ADS does not
currently work }

Shears J., Lloyd C., 2007, baavss-alert 1494 \\
\OEJVlink{http://tech.groups.yahoo.com/group/baavss-alert/message/1494}

Shears J., Lloyd C., Boyd D., Brady S., Miller I., Pickard R., 2008,
J. Brit. Astron. Assoc., in press,
\OEJVlink{http://adsabs.harvard.edu/abs/2008arXiv0805.1591S}

Vanmunster T., Fried R., 1999, vsnet-alert 3707 (see Appendix 3)

\end{document}